# Experimental study of the behavioural mechanisms underlying self-organization in human crowds

Mehdi Moussaïd[1, 2, 3], Dirk Helbing[1], Simon Garnier[2], Anders Johansson[1], Maud Combe[2] and Guy Theraulaz[2]

[1] ETH Zurich, Swiss Federal Institute of Technology, Chair of Sociology, UNO D11, Universitätstrasse 41, 8092 Zurich, Switzerland. Phone +41 44 632 88 81; Fax +41 44 632 17 67.

[2] Centre de Recherches sur la Cognition Animale, UMR-CNRS 5169, Université Paul Sabatier, Bât 4R3, 118 Route de Narbonne, 31062 Toulouse cedex 9, France. Phone: +33 5 61 55 64 41 ; Fax: +33 5 61 55 61 54.

[3] Corresponding author: moussaid@cict.fr

## Abstract

In animal societies as well as in human crowds, many observed collective behaviours result from self-organized processes based on local interactions among individuals. However, models of crowd dynamics are still lacking a systematic individual-level experimental verification, and the local mechanisms underlying the formation of collective patterns are not yet known in detail. We have conducted a set of well-controlled experiments with pedestrians performing simple avoidance tasks in order to determine the laws ruling their behaviour during interactions. The analysis of the large trajectory dataset was used to compute a behavioural map that describes the average change of the direction and speed of a pedestrian for various interaction distances and angles. The experimental results reveal features of the decision process when pedestrians choose the side on which they evade, and show a side preference that is amplified by mutual interactions.

The predictions of a binary interaction model based on the above findings were then compared to bidirectional flows of people recorded in a crowded street. Simulations generate two asymmetric lanes with opposite directions of motion, in quantitative agreement with our empirical observations. The knowledge of pedestrian behavioural laws is an important step ahead in the understanding of the underlying dynamics of crowd behaviour and allows for reliable predictions of collective pedestrian movements under natural conditions.

**Keywords**: Self-organization – Crowds – Pedestrian interactions – Social force model – Controlled experiments

## *1. Introduction*

The idea that large-scale collective behaviour emerges from local interactions among individuals has become a key concept in the understanding of human crowd dynamics (Ball 2004; Couzin & Krause 2003; Helbing et al. 2001; Sumpter 2006). Examples of such collective behaviours are the spontaneous formation of lanes of uniform walking direction in bidirectional flows (Milgram & Toch 1969) or the oscillation of the passing direction at narrow bottlenecks (Helbing & Molnar 1995). The quantitative understanding of these collective phenomena is a major precondition for the prediction of congestion, the planning of evacuation strategies, and the assessment of building or urban layouts. Therefore, recent research tries to understand how pedestrians move and interact with each other in order to predict the phenomena emerging at the scale of a crowd (Dyer et al. 2007; Helbing et al. 2000; Helbing et al. 1997; Yu & Johansson 2007).

Many models of pedestrian behaviour have been suggested to describe the mechanisms leading to the formation of collective patterns (Antonini et al. 2006; Burstedde et al. 2001; Kirchner & Schadschneider 2002; Willis et al. 2000). In particular, the use of attraction and repulsion forces to describe the motion of a pedestrian has generated promising results (Helbing 1991; Hoogendoorn & Bovy 2003; Yu et al. 2005). For instance, the *social force model* has been successful in qualitatively reproducing various observed phenomena and has been adapted many times when addressing problems of crowd modelling (Helbing & Molnar 1995; Johansson et al. 2007; Lakoba et al. 2005). The basic modelling concept suggests that the motion of a pedestrian can be described by the combination of a driving force, that reflects the pedestrian's internal motivation to move in a given direction at a certain desired speed, and repulsive forces describing the effects of interactions with other pedestrians and boundaries such as walls or obstacles in streets.

However, the underlying assumptions and the exact form of the forces involved have never been empirically measured or validated, although the function describing the interactions among individuals is likely to play a significant role for the resulting collective patterns, as it has been demonstrated for various social animal species (Couzin et al. 2002; Dussutour et al. 2005). The most accurate studies, so far, were restricted to calibrating parameters of assumed interaction forces by minimizing the error in predicting individual motion (Hoogendoorn & Daamen 2007; Johansson et al. 2007).

In this study, we use a novel approach by measuring the behavioural effects of interactions between pedestrians in controlled experiments. Indeed, under controlled conditions, the response of individuals to mutual interactions can be easily observed and described in statistical terms, which then enables the extraction of individual behavioural laws. Similar experimental approaches were successfully applied in the past to grasp the behavioural mechanisms underlying the self-organization in many social animal species (Ame et al. 2006; Beekman et al. 2001; Buhl et al. 2006; Camazine et al. 2001; Dussutour et al. 2004; Jeanson et al. 2005; Millor et al. 1999; Theraulaz et al. 2002; Ward et al. 2008).

Considerable progress in tracking technologies has made such an approach possible for the study of crowd dynamics, and has recently motivated a series of experiments on crowds: One the one hand, various studies have aimed at characterizing *macroscopic* crowd patterns, such as the speed-density diagram (Seyfried et al. 2005), the flow around a bottleneck (Helbing et al. 2005; Hoogendoorn & Daamen 2005; Kretz et al. 2006), or the collective dynamics during evacuation processes (Isobe et al. 2004). One the other hand, several studies have investigated various aspects of *microscopic* pedestrian motion, such as step frequencies (Hoogendoorn & Daamen 2005) or lateral body oscillations (Fruin 1971; Pauls et al. 2007). Our experimental approach, in contrast, aims at linking both the individual and collective level of observations by measuring the interaction laws between individuals. How does a pedestrian modify the behaviour in response to interactions with other pedestrians? Answering this question can reveal the precise mechanisms leading to the self-organization in crowds and help to construct reliable crowd models.

To tackle this question, we have observed the behaviour of a pedestrian moving in a corridor under three different experimental conditions: (1) in the absence of interactions, (2) in response to a standing pedestrian, and (3) in response to a pedestrian moving in opposite direction. The comparison of pedestrians trajectories with and without interactions allowed us to quantify the behavioural effects of interactions. The laws describing the interactions were then formalized in mathematical terms and implemented in the social force model. We finally compared the predictions of the model with the experimental results and empirical data of pedestrian flows recorded in a crowded street.

## 2. Material and Methods

### (a) Laboratory experiments

Controlled experiments have been conducted from February to March 2006 at the Hospital Pellegrin, in Bordeaux (France). Twenty females and twenty males aged 18 to 30 and naïve to the purpose of the experiment agreed to participate in the study and gave informed consent to the experimental procedure. The study has been approved by the Ethics Committee of the *Centre Hospitalier Universitaire de Bordeaux*. The experimental corridor (Length=7.88m, Width=1.75m) was equipped with a tracking system, which consisted of three digital cameras (SONY DCR-TRV950E, 720x576 pixels) mounted at the corners of the corridor (Figure 1). Participants were equipped with a white T-shirt and coloured table tennis balls on their shoulders to facilitate an accurate detection of their motion by cameras. The 3D reconstruction of the shoulder position was made on the basis of the digital movies of all three cameras, encoded at 12 frames per second, and with the help of a specialized software developed in accordance with the procedures described in Ref. (Bouguet). Cameras were calibrated in space by using a planar checkerboard, and in time by switching off and on the light at the beginning of the recordings. The 3D data were finally projected to the 2D floor, and each pedestrian was characterized by a single point located at the middle of the line connecting both shoulders positions. The trajectories were finally smoothed over a time window of 10 frames.

The forty selected subjects were divided into 8 groups of 5 people each. One session was performed on each day. Every session was carried out with one of the 8 groups constituted before. It was randomly chosen and participated only once. During each session, five replications of the following conditions were performed: (1) In condition 1, a single pedestrian was given the instruction to go back and forth in the corridor during a period of 3 minutes (which corresponds to approximately 20 trajectories, 10 in each direction). Every subject performed this condition once. (2) In condition 2, one subject was instructed to stand still in the middle of the corridor, while another one received the same instructions as in condition 1, and therefore had to evade the standing pedestrian. Each participant performed the test once as a walker and once as an ”obstacle”. Each replication lasted for 3 minutes. (3) In condition 3, two subjects received the same instructions as in condition 1, but starting from opposite ends of the corridor, and therefore had to evade each other. A starting signal was given for each new trial, so that pedestrians always met each other in the center of the corridor. Pairs of participants were chosen randomly, and each replication ended after 20 trials. We have reconstructed 90, 148 and 123 trajectories for conditions 1, 2 and 3 respectively. For a better fit of the acceleration behaviour, the data from condition 1 were complemented with additional data obtained under the same conditions, but with a more accurate tracking system (Vicon Motion tracking system).

### (b) Field observations

Bidirectional flows of pedestrians were observed in a pedestrian zone in Bordeaux, France (Sainte-Catherine street, during April 2007). The street was video-recorded from above with a digital camera (SONY DCR-TRV950E, 720x576 pixels) during 30 minutes and at a height of approximately 10 meters. The picture field covered an area of 8x6 meters. A 1 meter wide area on each side of the street was occupied by a few pedestrians standing next to shops, while the flow of walkers was mainly concentrated in the middle of the street. The lens distortion was corrected, and 2670 pedestrians were tracked at a rate of 1 frame per second.

### (c) Simulation design

Simulations were performed in a way reflecting the experimental conditions, with simulated pedestrians starting from a 20cm squared area located at each end of the corridor. The parameters used were $v0 = 1.3$m/s, $\tau = 0.5$s, and the respective destinations were assumed to be located 0.5m after the end of the corridor, to allow for some flexibility towards the end of the trajectory. The time step was set to $\delta t = 1/20$s.

Simulations of bidirectional flows were conducted in a 6x50m street with the pedestrians of each flow starting from the central, 4m wide area (i.e. area 2, 3, 4 and 5 in Figure 5a). As observed in field observations, simulated pedestrians entered the street at a rate of 0.65 per second, with an initial speed $v = v0 = 1.2 \pm 0.4$m/s. Borders of the street (1m on each side) were occupied by randomly located static pedestrians with a density of 0.2p/m$^2$. The results shown Figure 5 were obtained by evaluation of 10 simulation runs over 10 minutes each to reflect the observed conditions.

## *3. Model description*

In accordance with the social force concept (Helbing & Molnar 1995), we consider that the motion of a pedestrian i can be described by means of three different components: (1) the internal acceleration behaviour $\vec{f}_i^{\,0}$, reflecting the pedestrian's motivation to move in a particular direction at a certain speed, (2) the effects of corridor walls $\vec{f}_i^{\,wall}$ on this pedestrian, and (3) the interaction effects $\vec{f}_{ij}$, reflecting the response of pedestrian i to another pedestrian j.

At a given moment of time, the change of velocity $\vec{v}_i$ of pedestrian i is then given by the equation:

$$\frac{d\vec{v}_i}{dt} = \vec{f}_i^{\,0} + \vec{f}_i^{\,wall} + \vec{f}_{ij}$$

In the following, we use experimental data to check the validity of the above equation and determine the interaction function $\vec{f}_{ij}$.

## *4. Measurement of the behavioural laws*

### 4.1 Single pedestrian behaviour

The experimental condition 1 was used to validate and calibrate the internal acceleration behaviour. Helbing and Molnar (Helbing & Molnar 1995) suggested the equation

$$\vec{f}_i^{\,0} = \frac{d\vec{v}_i}{dt} = \frac{v_i^0 \vec{e}_i^{\,0} - \vec{v}_i(t)}{\tau}$$

describing the adaptation of the current velocity $\vec{v}_i$ of pedestrian i to a desired speed $v_i^0$ and a desired direction of motion $\vec{e}_i^{\,0}$ (given by the direction of the corridor) within a certain relaxation time $\tau$ . According to Figure 1, this equation describes the observed acceleration behaviour well. The desired velocities $v_i^0$ are normally distributed with an average value of 1.29±0.19m/s (mean±sd), and the relaxation time amounts to $\tau = 0.54 \pm 0.05$ seconds (see Figure 1b).

### 4.2 Interaction law

Conditions 2 and 3 were then used to determine the exact trajectories when avoiding a standing or moving pedestrian (Figure 2). By the formula

$$\vec{f}_{ij}(t) = \frac{d\vec{v}_i}{dt} - \vec{f}_i^{\,0}(t) - \vec{f}_i^{\,wall}(t)$$

we have measured the interaction effect $\vec{f}_{ij}$ resulting from the interaction with the other pedestrian *j*. In the above equation, the term $\vec{f}_i^{\,0}$ has been calibrated during the experimental condition 1, while the interactions with the corridor walls $\vec{f}_i^{\,wall}$ have been specified according to previous findings (Johansson et al. 2007) that is, as a function of the distance $d_w$ perpendicular to the wall: $\vec{f}_i^{\,wall}(d_w) = ae^{-d_w/b}$, with parameters a = 3 and b = 0.1 corresponding to a repulsion strength of the same order as the internal acceleration term, and a repulsion range of approximately 30cm from the wall border.

We then quantified the interaction laws of pedestrians by computing the average value of the interaction effect $\vec{f}_{ij}(t)$ at different interaction distances and angles. For this, we partitioned the area in front of pedestrian *i* into a 15x25 grid. In each cell of the grid, we computed the mean interaction effect $\langle \vec{f}_{ij} \rangle$ resulting from the presence of pedestrian *j* in this cell, averaged over all the trajectories of the experimental condition 2 (N=148) (see Figure 6 of the Supplementary Material). This finally provides us with a so-called behavioural map, which summarizes the average change of speed and direction of the focal pedestrian *i* in various interaction configurations (Figure 3).

It was not obvious in advance that removing the effects of internal acceleration and walls would yield a highly structured vector field, which can be interpreted as the outcome of characteristic interpersonal interactions. However, the resulting values of $\langle \vec{f}_{ij} \rangle$ as a function of the distance and the angle of approach turn out to show a clear and reasonable dependence. In contrast to previous heuristic specifications, we find that a pedestrian *i* essentially continues to move at the previous speed and mainly adjusts the direction of walking, when another pedestrian *j* is located towards the sides (i.e. either the left-hand side when x<-0.25m or the right-hand side when x>0.25m). Pedestrians decelerate significantly primarily in case of head-on encounters, i.e. when the pedestrian *j* is located in front of pedestrian *i* (see the light grey area in figure 3). This corresponds to the zone where pedestrians choose the side on which they want to pass. For this reason, we interpret this central area as a *decision zone*: for head-on encounters, it is necessary to take a binary decision, whether to evade the other pedestrian on the left-hand side or on the right-hand side. Moreover, it turns out that the resulting choice of the passing side is biased. Pedestrians avoiding a static pedestrian have a slight preference for the right-hand side in our experiments, but the asymmetry is significantly more pronounced if both pedestrians are moving (see blue bars in Figure 4). This shows that the mutual adjustment of motion of two interacting pedestrians amplifies the individual left/right bias significantly.

## 4.3 Specification of the interaction laws

Given the above experimental observations, we now model the interaction function $\vec{f}_{ij}$ by fitting the extracted behavioural map. In the previous section, we have described the interaction effects in terms of directional changes (towards the sides) and speed changes (during head-on encounters). Therefore, it is natural to specify the interaction function on the basis of two components, $f_v$ and $f_\theta$, describing the *deceleration* along the interaction direction $\vec{t}_{ij}$ and *directional changes* along $\vec{n}_{ij}$ respectively, where $\vec{n}_{ij}$ is the normal vector of $\vec{t}_{ij}$, oriented to the left (see, for example, Hoogendoorn & Bovy, 2003 for a similar representation).

We specify the interaction direction $\vec{t}_{ij}$ as a composition of the direction of relative motion $(\vec{v}_i - \vec{v}_j)$ and the direction $\vec{e}_{ij} = (\vec{x}_j - \vec{x}_i) / \|\vec{x}_j - \vec{x}_i\|$, in which pedestrian j is located, where $\vec{x}_i$ is the location of pedestrian i. This leads to $\vec{t}_{ij} = \vec{D}_{ij} / \|\vec{D}_{ij}\|$ with $\vec{D}_{ij} = \lambda(\vec{v}_i - \vec{v}_i) + \vec{e}_{ij}$, where the weight λ reflects the relative importance of the two directions. The value estimated from the experimental data is $\lambda = 2.0 \pm 0.2$.

If $d_{ij}$ denotes the distance between two pedestrians *i* and *j*, and $\theta_{ij}$ the angle between the interaction direction $\vec{t}_{ij}$ and the vector pointing from pedestrian *i* to *j*, fitting our experimental data yields the following mathematical functions:

$$f_v(d,\theta) = -Ae^{-d/B-(n'B\theta)^2} \qquad (4)$$

and

$$f_\theta(d,\theta) = -AKe^{-d/B-(nB\theta)^2} \qquad (5)$$

(see Figure 3). There, we have dropped the indices *i* and *j*. K = θ/|θ| is the sign of the angle θ, and *A, B, n, n'* are model parameters. Equation (4) represents an exponential decay of the deceleration with distance d. The decay is faster for large values of θ, i.e. towards the sides of the pedestrian. Therefore, the deceleration effect is strongest in front, in accordance with the decision area identified above. Through the dependence on $B = \gamma\|\vec{D}\|$, it is increased in the interaction direction by large relative speeds, while the repulsion towards the sides is reduced. This reflects the fact that fast relative motions require evading decisions in a larger distance, which also means that the same amount of displacement to the side (basically the shoulder width plus some safety distance) can be gained over a longer way, requiring a weaker sideward movement (compare Figure 2a with 2b). Note that Eq. (5) is analogous to Eq. (4) for the directional changes, just with another parameter $n<n'$, which corresponds to a larger angular interaction range. The prefactor $K = \theta/|\theta|$ takes into account the discontinuity in the angular motion, reflecting the binary decision to evade the other pedestrian either to the left or to the right.

The resulting interaction effect $\vec{f}_{ij}$ becomes clearer, if we sum up over all contributing terms, resulting in

$$\vec{f}_{ij}(d,\theta) = -Ae^{-d/B}\left[e^{-(n'B\theta)^2}\vec{t} + e^{-(nB\theta)^2}\vec{n}\right] . \qquad (6)$$

Accordingly, we have an exponential decay of the interaction effect with the pedestrian distance *d*, where the interaction range *B* depends on the relative speed. The angular dependence and anisotropy of the interactions is reflected by the *θ*-dependence. The model parameters have been estimated from the experimental data to be A = 4.5±0.3, γ = 0.35±0.01, n = 2.0±0.1 and n = 3.0±0.7, by using of an evolutionary algorithm designed to minimize the difference between observed and simulated trajectories in conditions 2 and 3.

Finally, the model has to take into account the observed asymmetry in the avoidance behaviour, which is reflected by the somewhat higher proportion of pedestrians evading on one side. The simplest way to reproduce this bias is to replace the angle θ in equation (6) by $\theta + B\varepsilon$ where $\varepsilon = 0.005 > 0$ corresponds to a preference for the right-hand side. The dependence on B describes the fact that pedestrians make a faster side choice when the relative speed increases. Note that in other countries like Japan, the pedestrians have a preference for the left-hand side (Helbing et al., 2005), which corresponds to a negative value of $\varepsilon$.

## 4.4 Comparison of model predictions with empirical results

*(i) Binary interactions*

After the above model was fitted to the experimental data, we have first tested it through a series of computer simulations involving two pedestrians in situations similar to conditions 2 and 3. The model predictions show that the shape of the trajectories during avoidance manoeuvres as well as the side choice proportions are in good agreement with the empirical data collected in our experiments (Figure 4). The non-trivial reinforcement of the side preference observed in conditions 2 and 3 is also well reproduced by the model. This first validation step demonstrates that the interaction function and the side preference have been well specified.

*(ii) Collective patterns*

We then used the model to study the dynamics of a larger number of pedestrians, who were exposed to many simultaneous interactions. In our simulation study, it was assumed that the behaviour of all pedestrians was simply given by the sum of all binary interactions with other pedestrians in the neighbourhood. The superposition of binary interaction effects was used to compare computer simulations of pedestrian counterflows with empirical data of collective pedestrian movements. For this, we conducted simulations of the above model under conditions reflecting the field observations (section 2.c). First, we found that restricting the number of neighbouring individuals a pedestrian responds to did not improve our results significantly. Therefore, the superposition of all binary interactions worked well for the above model. Second, we observed that the empirical flows, as well as the simulated ones, displayed two lanes of pedestrians moving in opposite direction. Moreover, both simulated and observed patterns exhibited a very pronounced left-right asymmetry in street usage (Figure 5b), while simulations for a uni-directional flow generate a uniform distribution of pedestrians (see the inset in Figure 5b). We also found an almost uniform distribution for bidirectional pedestrian flows of low density, which supports the idea that a minimum amount of interactions is necessary for the flow separation to emerge (see Figure 6 in Supp. Info.).

## *5. Discussion*

We have presented a set of controlled experiments that revealed the detailed mechanisms and functional dependencies of pedestrian interactions in space and time. In contrast to previous modelling approaches, we did not use a prefabricated interaction function and fitted parameters to the data. Instead, we first extracted dependencies between certain variables from the data (such as the longitudinal and the lateral movement components as a function the relative positions of interacting pedestrians). Then, we identified suitable mathematical functions fitting them. Only after such functions were extracted, the model parameters were determined. Therefore, the interaction function is not just chosen in a plausible way, but it explicitly represents experimentally determined features of the data.

Our experimental result reveals how pedestrians modify their behaviour during interactions. Towards the side of another pedestrian, people simply adjust their direction of motion to avoid collisions. In case of head-on encounters, a binary decision takes place: pedestrians need choose whether to evade the other person on the right-hand or on the left-hand side. This decision process goes along with a significant decrease of walking speed. During evading manoeuvres, there is an individual bias towards one side. This seems to make the movement smoother and to reduce the related speed decrease.

The side preference is not directly coupled to the asymmetry of the body, nor to the direction of car traffic, as can be illustrated, for example, by the observed right-hand traffic organization in some areas of Great Britain (Moussaid et al. 2009; Older 1968). Instead, we suggest that the left/right bias can be interpreted as a behavioural convention that emerges because the coordination during evading manoeuvres is enhanced when both pedestrians favour the same side (Bolay 1998; Helbing 1991). It is therefore advantageous for an individual to develop the same preference as the majority of people. Through a self-reinforcing process, most people would use the same strategy in the end. As both sides are equivalent in the beginning, the theory predicts that different preferences emerge in different regions of the world, as it is actually observed (Helbing et al. 2001).

In other words, the side preference may be interpreted as a cultural bias. This cultural interpretation could potentially be checked by performing walking experiments with young children, but these may involve a variety of ethical and organizational issues. Alternatively, the hypothesis that asymmetrical evading behaviour is based on a self-organized behavioural convention could also be tested by empirically studying the occurrence of a side preference in some hard-to-reach areas with high population densities, but no car traffic. Furthermore, other examples of "coordination games" could be investigated as well (e.g. regarding the writing direction, clock direction, side of hot water tap, VHS vs. BetaMax video format (Arthur 1990), DVD format, etc.). In such coordination games, symmetry breaking occurs in the initial phase of the self-organized formation of a convention (Helbing 1991; Helbing et al. 2001). Eventually, however, the asymmetry becomes institutionalized, i.e. it becomes a cultural bias transmitted from one person to another by imitation and learning. In such a way, the asymmetrical behaviour is culturally inherited, and symmetry breaking is not spontaneous anymore.

We show that the concept of social forces is applicable in principle for modelling the observed movements of pedestrians during our experiments, and that it facilitates a quantitative prediction of collective crowd patterns. In particular, the interaction function is well described by the combination of a deceleration effect with directional changes. The deceleration effect applies to head-on encounters, when a binary decision between the right-hand and the left-hand side must be made, while directional adjustments apply otherwise and afterwards. Moreover, a simple bias in the interaction angle influences the statistical properties of the resulting collective patterns of motion. It supports the formation of a small number of lanes at high pedestrian densities (typically two), i.e. it separates the opposite walking directions very effectively and minimizes the frequency of mutual obstructions.

Our results also show that the amplification of the side preference at the crowd level requires the combination of asymmetric behaviour with frequent interactions to quantitatively reproduce empirical data on side preference. The left/right bias is much more pronounced when people have to mutually adjust to each other (as in condition 3). This may also explain the unexpected observation of higher traffic efficiency in some situations of counterflows (Algadhi et al. 2002; Helbing et al. 2005). Similar amplification phenomena, where an individual preference is amplified by the action of many other individuals and shapes the collective organization, have been recently observed in various other group-living organisms (Ame et al. 2004; Bon et al. 2005; Jeanson et al. 2005). For example, a slight wall-following tendency in ants affects the colony choice of a path to a food source (Dussutour et al. 2005). Our results, therefore, show that similar mechanisms seem to guide the dynamics of human crowds.

These findings may be used to assess the suitability of pedestrian facilities and escape routes under various conditions, such as the movement of homogeneous as compared to multinational crowds with different side preferences (e.g. during international sports events). This could significantly affect the efficiency of pedestrian flows during mass events or the functionality of heavily frequented buildings such as railway stations, if not taken into account in the planning of events and the dimensioning of public spaces and facilities.

Finally, we highlight the fact that experimental methods of investigation previously applied to the study of animal collective behaviour can be successfully transferred to the study of human interaction laws and to collective phenomena emerging from them (Dyer et al. 2007; Dyer et al. 2008), even though the behavioural and cognitive complexity of human are greater. We also note that a multivariate linear regression approach would not be able to identify laws resulting in self-organized collective behaviours, as those require non-linear interactions. Therefore, it is necessary to quantitatively extract the nonlinear dependencies from the data. In a similar way, one may address a multitude of other problems like the simultaneous interaction with several other people, or communication and decision-making behaviours explaining self-organized phenomena ranging from collective attention (Wu & Huberman 2007) over collective opinion formation (Deffuant et al. 2001), up to social activity patterns (Barabàsi 2005).

## Acknowledgements

We thank Dr. Colette Fabrigoule and Pr. Jean-François Dartigues for their support that made the experimental procedure possible. We also thank Pier Zanone, Vincent Fourcassié, Christian Jost, Niriaska Perozo, Anne Grimal, Wenjian Yu, Jeanne Gouëllo and the members of the EMCC group in Toulouse for inspiring discussions. This study was supported by grants from the CNRS (Concerted Action: ”Complex Systems in Human and Social Sciences”) and the University Paul Sabatier (*Aides Ponctuelles de Coopération*). Mehdi Moussaïd is supported by a jointly financed doctoral-engineer fellowship from the ETH Zürich and the CNRS. Simon Garnier is supported by a research grant from the French Ministry of Education, Research and Technology.

### Figure caption

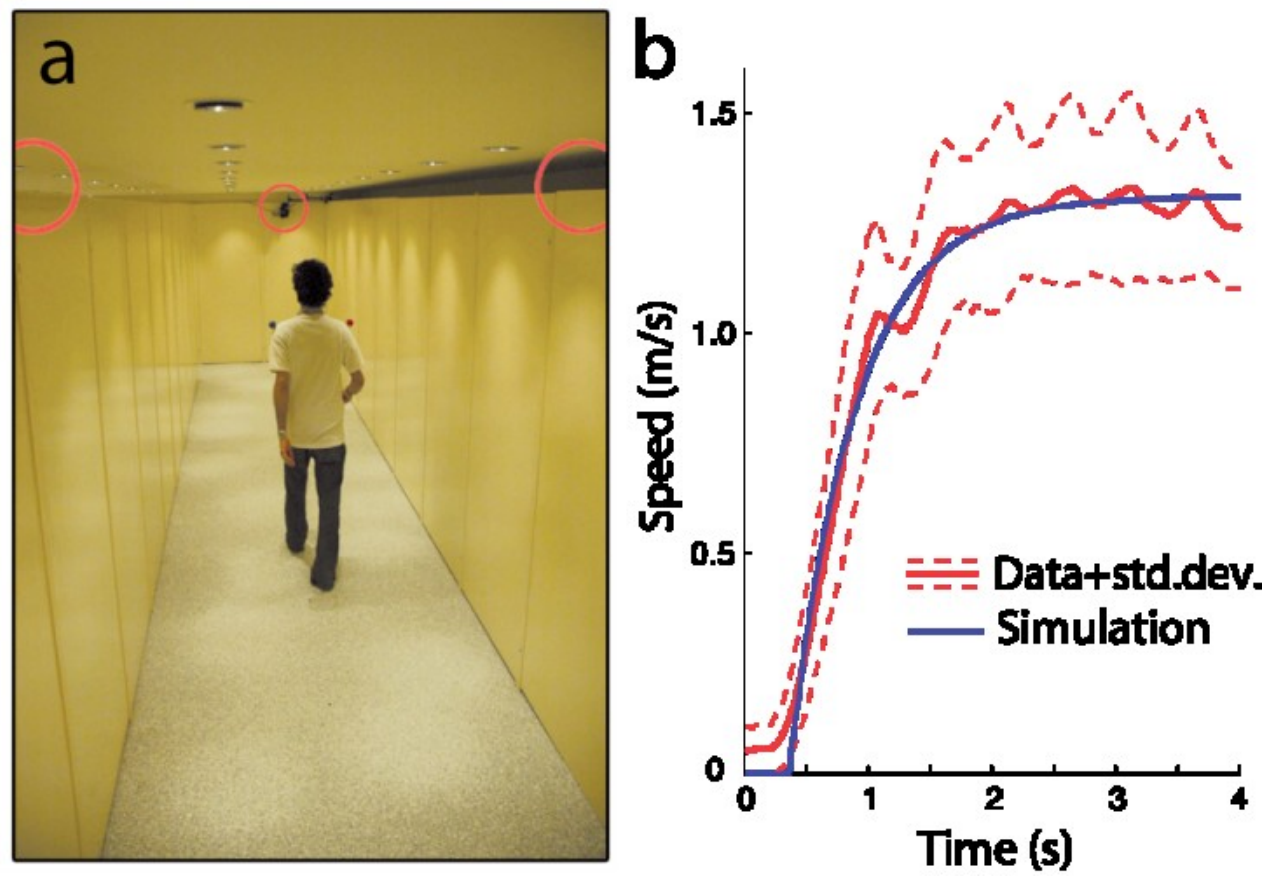

**Figure 1**: (a) Snapshot of the experimental setup. Red circles indicate the location of cameras. (b) Calibration of the acceleration behaviour on the basis of the average time-dependent pedestrian velocity in the absence of interactions. The fitted curve (blue) is given by the acceleration equation (2). The parameters were estimated as τ=0.54±0.05s and v0=1.29±0.19m/s after a reaction time of 0.35s.

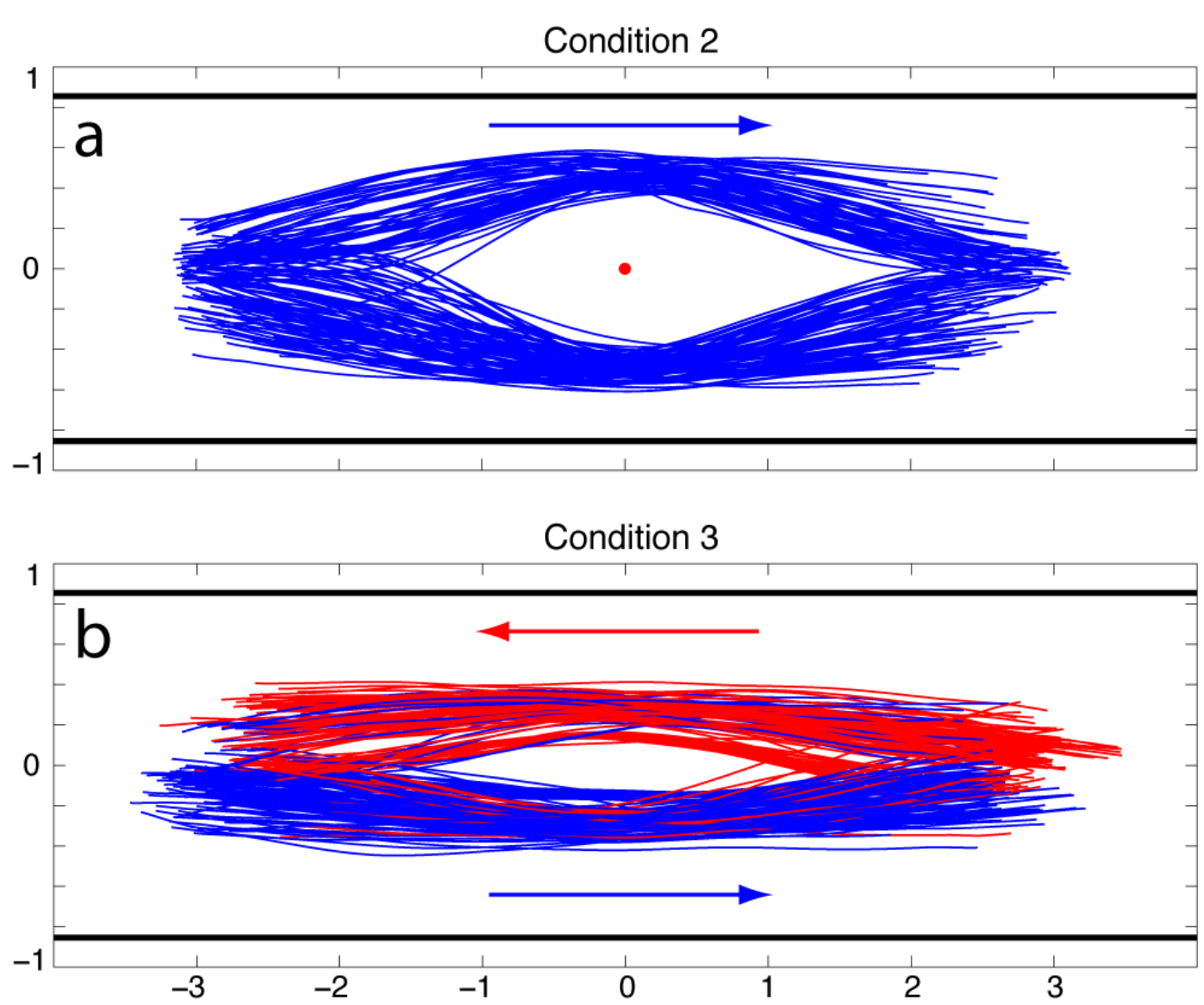

**Figure 2**: Observed trajectories in condition2 (N=148) and condition3 (N=123). One of the pedestrians (moving from left to right) is represented in blue, while the other one is represented in red.

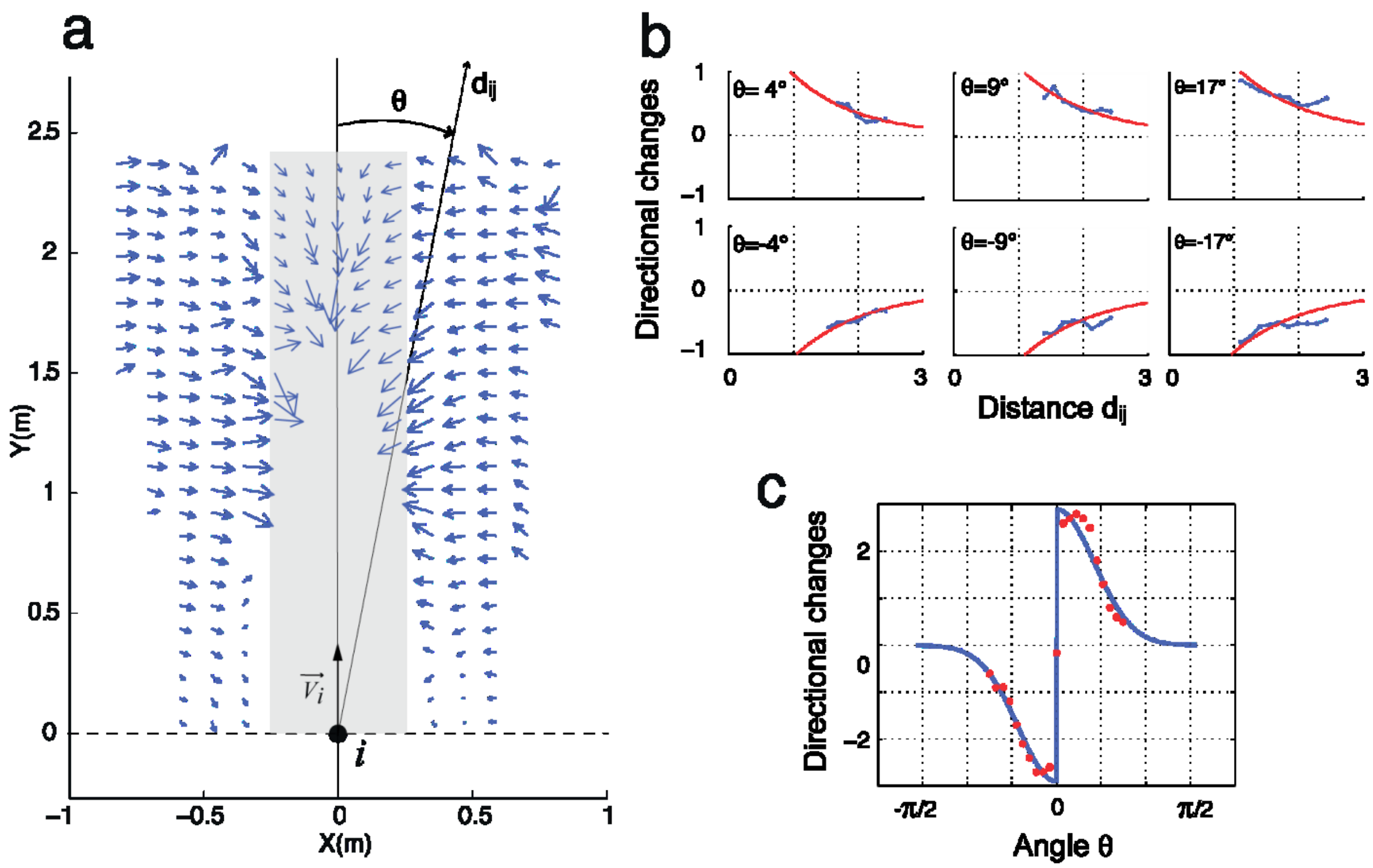

**Figure 3**: (a) Average value of the interaction effect $\vec{f}_{ij}$ at various distance $d_{ij}$ and angle $\theta_{ij}$ during experimental condition 2. (b) For a given angle $\theta$, the function $f_\theta(d,\theta)$ describing the directional changes, decreases exponentially with d, which provides the relation $f_\theta = A(\theta)e^{-bd}$, with fit parameter $b$. (c) $A(\theta)$ can then be approximated by the equation $aKe^{-(c\theta)^2}$, where K is the sign of $\theta$ and $a$, $c$ are fit parameters. The function $f_v(d,\theta)$ for speed changes has been set according to a similar functional dependency.

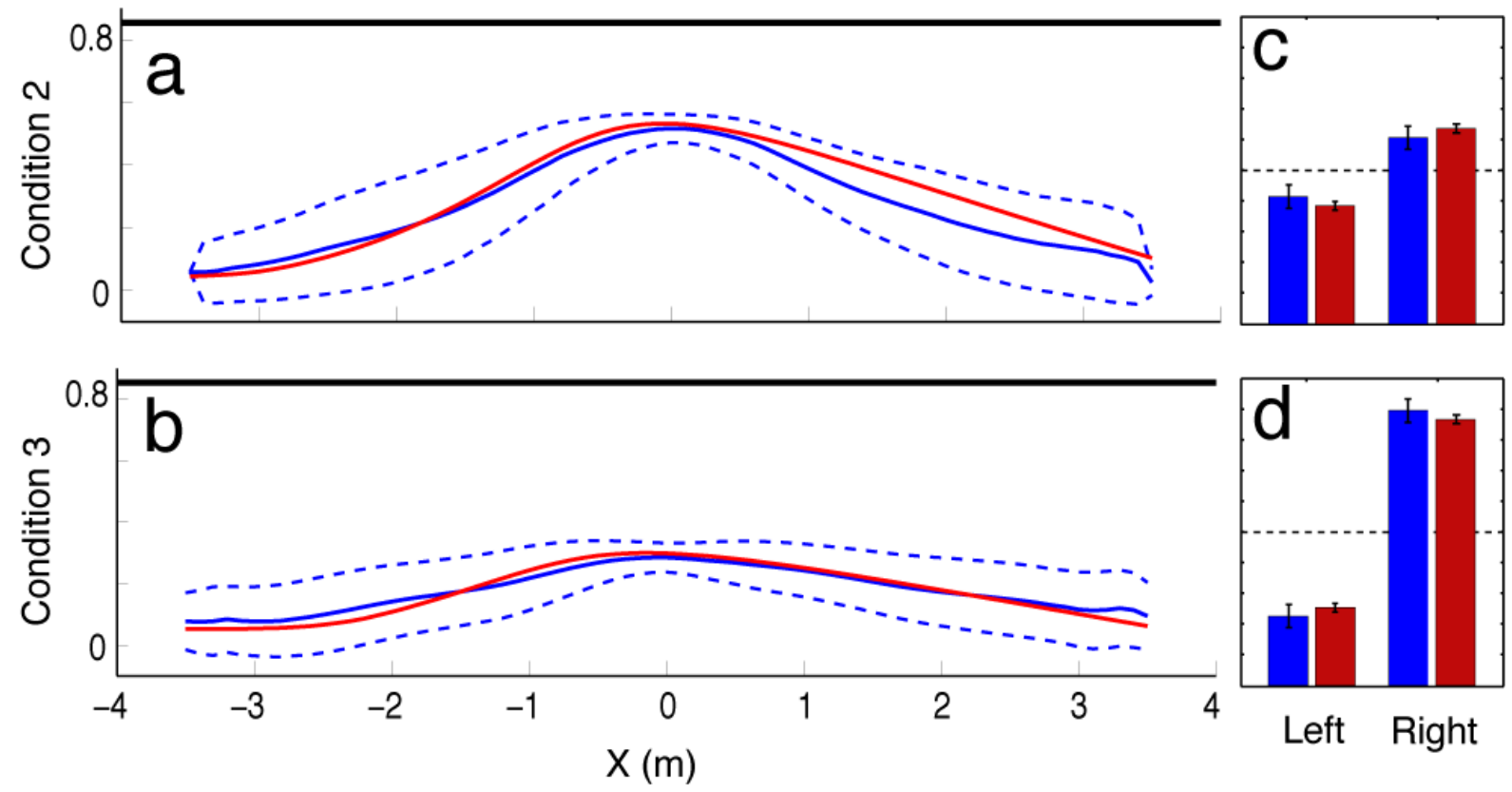

**Figure 4**: Numerical simulations as compared to experimental observations during conditions 2 and 3. In (a) and (b), the blue lines correspond to the average observed trajectories, with pedestrians moving from left to right. The blue dashed lines indicate the standard deviation. Red lines correspond to the average trajectories obtained after 1000 simulations (with parameter values $A$=4.5, $n$=2, $n'$=3 and =0.005). Bars in (c) and (d) indicate the proportions of choosing the left- or right-hand side in an avoidance manoeuvre during the experiment (blue) or in simulations (red).

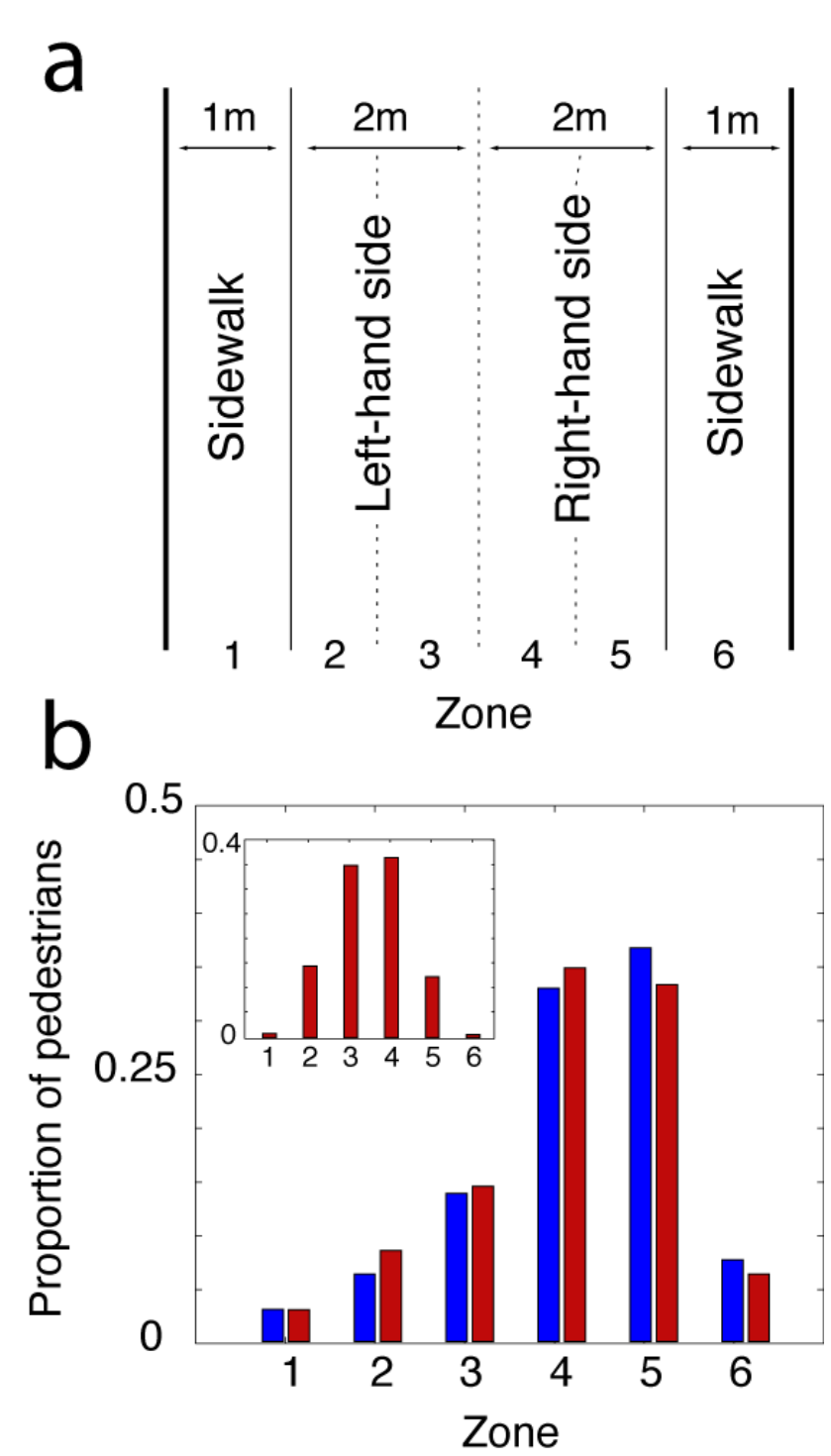

**Figure 5**: Asymmetry of bidirectional pedestrian traffic. As sketched in (a), six areas were distinguished for the measurements: 1) left sidewalk, 2) and 3) left side of the walkway, 4) and 5)

right side of the walkway and 6) right sidewalk. 'Left' and 'right' are referring to the walking direction. The sidewalks next to shops were occupied by a small number of standing pedestrians. The blue bars in (b) show the proportion of observed pedestrians walking in each area, while the red bars are simulation results (with the same parameter values as in Figure4). For comparison, the inset illustrates the symmetric simulation results for a unidirectional flow.